\documentclass[12pt]{article}

\pdfoutput=1

\usepackage[left=0.9in,top=0.9in,right=0.9in,bottom=0.9in]{geometry}

\usepackage{latexsym}
\usepackage{epsfig,amssymb,euscript}
\usepackage{amsmath}
\usepackage{mathrsfs}
\usepackage{ccaption}    
\usepackage[usenames,dvipsnames]{color}

\usepackage{ dsfont }

\usepackage{tikz}
\usetikzlibrary{positioning}

\usepackage[
      colorlinks=true,
      linkcolor=blue,
      urlcolor=blue,
      filecolor=black,
      citecolor=red,
      pdfstartview=FitV,
      pdftitle={},
        pdfauthor={Simon A. Gentle,Stefan Vandoren},
        pdfsubject={},
        pdfkeywords={},
        pdfpagemode={},
        bookmarksopen=true
      ]{hyperref}

  \captionnamefont{\bfseries}

\usepackage{subfigure}

\def\be{\begin{equation}}
\def\ee{\end{equation}}
\def\bea{\begin{eqnarray}}
\def\eea{\end{eqnarray}}

\usepackage{sectsty}
\sectionfont{\large}

\numberwithin{equation}{section}

\title{\bf \Large Lifshitz entanglement entropy from holographic cMERA}
\author{\large Simon A.~Gentle$^{a,b}$ and Stefan Vandoren$^a$\\ \\
        \small $^a$\it Institute for Theoretical Physics and Centre for Extreme Matter and Emergent Phenomena, \\
        \small \it Utrecht University, 3508TD Utrecht, The Netherlands \\ \\
        \small $^b$\it Instituut-Lorentz for Theoretical Physics, \\ 
        \small \it Leiden University,  2333CA Leiden, The Netherlands \\ \\
       \normalsize\href{mailto:s.a.gentle@uu.nl}{\texttt{s.a.gentle@uu.nl}}\texttt{, }\href{mailto:s.j.g.vandoren@uu.nl}{\texttt{s.j.g.vandoren@uu.nl}}}
\date{}

\begin{document}

\setlength{\baselineskip}{18pt}

\maketitle

\begin{abstract}
\setlength{\baselineskip}{18pt}
We study  entanglement entropy in free Lifshitz scalar field theories holographically by employing the metrics proposed by Nozaki, Ryu and Takayanagi in~\cite{Nozaki:2012zj} obtained from a continuous multi-scale entanglement renormalisation ansatz (cMERA). In these geometries we compute the minimal surface areas  governing the entanglement entropy as  functions of the dynamical exponent $z$ and we exhibit a transition from an area law to a volume law analytically in the limit of large $z$.  We move on to explore the effects of a massive deformation, obtaining results for any $z$ in arbitrary dimension. We then trigger a renormalisation group flow between a Lifshitz theory and a conformal theory and observe a monotonic decrease in entanglement entropy along this flow.  We focus on strip regions but also consider a disc in the undeformed theory. 
 \end{abstract}





\section{Introduction}

Lifshitz theories are characterised by a scaling symmetry under which space and time scale differently: 
\begin{equation}
\vec{x} \to \lambda \vec{x}, \quad t \to\lambda^z t \ ,
\end{equation}
where $z$ is referred to as the dynamical exponent.  Such theories govern quantum critical points in many condensed matter systems.  As a simple example, consider the following theory of a free massless scalar in $d+1$ dimensions:
\begin{equation}\label{eq:LifshitzTheory}
I=\frac{1}{2} \int {\rm d}^{d+1}x\left[(\partial_t\phi)^2-\alpha^2(\vec{\nabla}^z\phi)^2\right]\ .
\end{equation}

Little is known about entanglement entropy in such theories.   For the special case of $d=z=2$ one can map the ground state to a Euclidean conformal field theory in $1+1$ dimensions and compute some subleading universal terms in the entanglement entropy~\cite{Fradkin:2006mb,Hsu:2008af,Fradkin:2009dus,Zhou:2016ykv,Parker:2017lnh,Chen:2016kjp}, see also \cite{PhysRevB.80.184421,Oshikawa:2010kv}.  Other work on $z=2$ in $d=1$ can be found in \cite{Chen:2017txi,Chen:2017tij}. This will be useful for us to fix the overall normalization factor in the entanglement entropy obtained from cMERA for the case $z=2$, as we discuss in Section 3.1.

The discretised theory with $d=1$ and arbitrary $z$, including a mass term, was studied recently in~\cite{MohammadiMozaffar:2017nri,He:2017wla}  and some partial results for $d=2$ were obtained in \cite{MohammadiMozaffar:2017nri}. In \cite{He:2017wla} an analytical approach was also put forward based on the holographic cMERA technique \cite{Nozaki:2012zj} for Lifshitz scalar fields. In this paper, we follow up and elaborate on this discussion and extend it to higher dimensions, obtaining several new results.

Holography opens up a new way to compute entanglement entropy. The difficult direct field theory calculation is mapped to a geometric extremisation problem in the dual gravity theory via the Ryu-Takayangi (RT) prescription \cite{Ryu:2006bv,Ryu:2006ef}.  In detail, first one computes the bulk metric that describes the field theory state of interest.  The entanglement entropy of a given region is then equal to the area of the minimal area extremal surface, as measured by this metric, that ends on this  region at the boundary. 

However, the holographic dual of a Lifshitz theory has yet to be universally agreed upon.  Whilst a dual spacetime, termed Lifshitz spacetime, was proposed in \cite{Kachru:2008yh,Taylor:2008tg} and has been studied intensively ever since (see \cite{Taylor:2015glc} for a  review), it is unclear whether the RT prescription should be applied to this spacetime.   Indeed, a study of various perspectives on the holographic reconstruction of Lifshitz spacetime can be found in \cite{Gentle:2015cfp}.  Other recent work suggests that Newton-Cartan geometry may provide a more natural bulk dual for a non-relativistic theory~\cite{Hartong:2014pma}. Regardless, holography typically computes the entanglement entropy for strongly-coupled field theories with large central charges, whereas here we focus on a very different setting.

In this paper we use a method inspired by holography that is applicable to free field theories. In particular,  Nozaki-Ryu-Takayanagi (NRT) proposed in \cite{Nozaki:2012zj} that a metric emerges from a continuous version of the multi-scale entanglement renormalisation ansatz  (cMERA) ~\cite{Haegeman:2011uy}. For a given theory, expectation values of the appropriate disentangler operator determine various components of this metric.   In some sense, the NRT proposal geometrises the entanglement entropy of free fields.  Our goal is to compute entanglement entropies by applying the RT prescription to the cMERA metric for various Lifshitz theories.  As we explain later, our results should be viewed as predictions for field theory calculations. We should stress however that, while the NRT proposal is holography inspired, it is not embedded within the AdS/CFT correspondence, since we apply it to a single free scalar field that is neither large $N$ nor at strong coupling. It is well known that free conformal fields do not have a gravity dual, at least not with a weakly coupled gravity sector coupled to matter fields. Nevertheless, the NRT proposal is similar in flavor to the RT prescription at a technical level (extremisation of surface areas using metric geometries), so we can make concrete calculations. The justification of this comes from the MERA approach, and the expectation that a continuum version of it should exist.

  We begin in the following section with a brief review of cMERA and the definition of the cMERA metric.  As a crucial consistency check we first compute  entanglement entropy in the ground state of a relativisitic free massive scalar theory and compare with known results.  We then turn to our three main calculations of Lifshitz entanglement entropy in Section~\ref{sec:Lifshitz}. We consider the original Lifshitz theory \eqref{eq:LifshitzTheory} and two  relevant deformations: a mass term $m^2\phi^2$ and a relativistic term $(\vec{\nabla}\phi)^2$.  We conclude with a discussion in Section~\ref{sec:Discussion}.

\section{cMERA and holographic entanglement entropy}

The multi-scale entanglement renormalisation ansatz (MERA) is a variational approach based on the renormalisation group (RG) to construct (ground) states in quantum many body systems and study their entanglement properties \cite{Vidal:2007hda,Vidal:2008zz}.  A continuum version was developed in~\cite{Haegeman:2011uy} for free fields, which we now summarise.  We follow the notation of~\cite{Nozaki:2012zj}.

Choose a quantum field theory in $d+1$ dimensions and introduce a length cut-off~$\varepsilon$. Consider then a one-parameter family of states  $|\Psi(u)\rangle$  living in the Hilbert space of this theory. The dimensionless parameter $u$ keeps track of the current length scale, with $u_{\text{UV}}=0$ and $u_{\text{IR}}=-\infty$ in the ultraviolet and infrared, respectively.  Focus initially on a reference state $ |\Psi(u_{\textrm{IR}})\rangle$ that has no entanglement. Run up the RG scale and generate entanglement by acting with a unitary transformation based on a local operator ${\cal K}(u)$.  Next act with a scale transformation ${\cal L}$ to introduce new degrees of freedom at shorter length scales.   Repeat this process  until the UV cut-off is reached.  The final  state  we are interested in can then be written as
\begin{equation}
 |\Psi(u_{\textrm{UV}})\rangle = U(u_{\textrm{UV}},u_{\textrm{IR}})|\Psi(u_{\textrm{IR}})\rangle, \quad U(u_1,u_2) \equiv P \exp\left[-i\int_{u_2}^{u_1}du\left({\cal K}(u)+{\cal L}\right)\right]
\end{equation}
and the variational principle can then be applied to minimise the energy of this state.  The variational parameters are encoded in the coefficients of the interactions in ${\cal K}(u)$.  Note that the states $|\Psi(u)\rangle$  are manifestly translationally invariant if these coefficients are independent of~$\vec{x}$.

\subsection{Metrics from cMERA}

The authors of~\cite{Nozaki:2012zj} associate a metric in $d+2$ dimensions with a type of cMERA.  In particular, they argue that a cMERA yielding a translationally invariant ground state should correspond to a metric of the form
\begin{equation}\label{eq:NRTMetric} 
{\rm d}s^2 = g_{uu}(u)\, {\rm d}u^2 + \frac{e^{2u}}{\varepsilon^2}\, {\rm d}\vec{x}^{\,2}_d+g_{tt}(u)\, {\rm d}t^2 \ .
\end{equation}
The metric in the RG direction parametrised by $u$ is given by the Hilbert-Schmidt distance between cMERA states at nearby scales $u$.  It can be expressed in terms of the variance of the operator ${\cal K}(u)$ in the state $|\Psi(u)\rangle$\footnote{Equation \eqref{eq:NRTguu} is a simple rewrite of equation (90) in \cite{Nozaki:2012zj} using their equations (18) and (21), relating $|\Psi(u)\rangle$ and ${\cal K}(u)$ with $|\Phi(u)\rangle$ and ${\hat{\cal K}}(u)$.} :
\begin{equation}\label{eq:NRTguu}
g_{uu}(u)=\langle\Psi(u)|{\cal K}(u)^2|\Psi(u)\rangle - \langle\Psi(u)|{\cal K}(u)|\Psi(u)\rangle^2 \ . 
\end{equation}
This metric component effectively measures the density of disentanglers at the scale $u$. The $g_{tt}$ component cannot be determined from the field theory entanglement on a fixed-time slice and therefore plays no role in our discussion. 

The explicit form of this metric can be calculated for the ground state of a free scalar theory~\cite{Nozaki:2012zj}, given the choice of disentangler proposed in \cite{Haegeman:2011uy}.  It depends purely on the dispersion relation $\omega(k)$, with $k\equiv|\vec{k}|$, and will be used throughout this paper:
\begin{equation}\label{eq:NRTguuFreeScalar}
\sqrt{g_{uu}(u)} =\left.\frac{k\,\partial_{k}\, \omega}{2\, \omega}\right|_{k=e^u/\varepsilon}\ .
\end{equation}

The cMERA metric constructed in this way captures how the quantum degrees of freedom in the field theory are entangled with each other at different RG scales. It has the flavour of holography, but the precise connection with AdS/CFT is not understood since we are neither at large $N$ nor at strong coupling and no expression for $g_{tt}$ is provided. For relativistic conformal field theories, the holographic cMERA approach gives the AdS metric \cite{Nozaki:2012zj}. And as we review below, applying the RT prescription to the cMERA metric for massive scalar fields yields the correct answers for the entanglement entropy when the correlation length is small. Based on this, we take a pragmatic approach and apply the holographic cMERA techniques to non-relativistic theories with Lifshitz scaling and with mass deformations corresponding to small correlation lengths. The method yields predictions for the Lifshitz entanglement entropy for general values of the dynamical exponent $z$. 

\subsection{Consistency checks in relativistic theories}\label{sec:Consistency}

Before embarking on our main calculations, in this section we illustrate the consistency of this approach with a simple example: a free massive scalar field in $1+1$ dimensions.  The action for this theory is (we absorb a factor $c^2/\hbar$ in $m$ such that $m$ has dimension of inverse seconds)
\begin{equation}\label{eq:CFTTheory}
I=\frac{1}{2} \int {\rm d}^{2}x\left[(\partial_t\phi)^2-c^2(\partial_x\phi)^2-m^2\phi^2\right]
\end{equation}
and the dispersion relation is $\omega(k)^2 =c^2 k^2+m^2$.  Computing $g_{uu}$ via \eqref{eq:NRTguuFreeScalar}, we should therefore associate the following metric with the ground state of this theory:
\begin{equation}\label{eq:CFTMetric}
{\rm d}s^2 = \left[\frac{e^{2u}}{2(e^{2u}+(m\varepsilon/c)^2)}\right]^2 {\rm d}u^2 + \frac{e^{2u}}{\varepsilon^2}\, {\rm d}x^2+g_{tt}\, {\rm d}t^2\ .
\end{equation}
For $m=0$ this reduces to the metric on AdS$_3$ in Poincar\'e coordinates with $z=\varepsilon e^{-u}$ and $g_{tt}=-c^2/z^2$, together with a simple rescaling of $x$ and $t$. The AdS radius in this normalisation is $R_{\text{AdS}}=2$, but one can rescale the overall metric to get any radius.

Our task now is to compute entanglement entropy from this metric using the RT prescription.  We will focus on an interval in the $x$-direction of width $\ell$ at $t=0$.  It is useful to define the dimensionless quantities
\begin{equation}\label{eq:J1andJ2Relativistic}
J_1\equiv \frac{m\varepsilon}{c}\quad \text{and} \quad J_2\equiv \frac{m\ell}{c}\ .
\end{equation}
We want to compute entanglement entropy as a function of $J_1$ and $J_2$.  We require $J_1<1$ and $J_2 > J_1 $  to ensure that the cut-off $\varepsilon$ is the smallest length scale in the theory. Furthermore, the regime $J_2<1$ means that the correlation length $\xi\equiv c/m$ is larger than the subsystem size~$\ell$, whereas $J_2>1$ means that the correlation length is smaller than the subsystem size. The entanglement entropy is known to be different in these two regimes \cite{Calabrese:2004eu} --- a result we now  re-derive.

First  we change coordinates:
\begin{equation}
e^{2u}=\frac{\varepsilon^2}{r^2}-J_1^2 \quad \Longrightarrow \quad {\rm d}s^2 =  \frac{{\rm d}r^2}{4r^2} + \left(\frac{\varepsilon^2}{r^2}-J_1^2\right)\frac{{\rm d}x^2}{\varepsilon^2}+g_{tt}\, {\rm d}t^2\ .
\end{equation}
The boundaries of the $r$ coordinate are fixed by the limits of the cMERA length scale $u$:
\begin{equation}\label{eq:CoordinateLimits}
r_\textrm{UV}(u=0)=\frac{\varepsilon}{\sqrt{1+J_1^2}}, \quad \quad r_\textrm{IR}(u\to -\infty)=\frac{\varepsilon}{J_1}=\frac{c}{m}=\xi\ .
\end{equation}
We seek geodesics of this metric parametrised by $x(r)$ that end at $x(r_\textrm{UV})=\pm\ell/2$.  The length of the shortest geodesic is proportional to the entanglement entropy. The steps to find the appropriate geodesics and compute their lengths are identical to those in \cite{Ryu:2006bv,Ryu:2006ef}; we adapt them to our setup and only present the results and a few intermediate steps.

\begin{figure}[!h]
\vskip0.5em
\begin{center}
\includegraphics[width=0.45\textwidth]{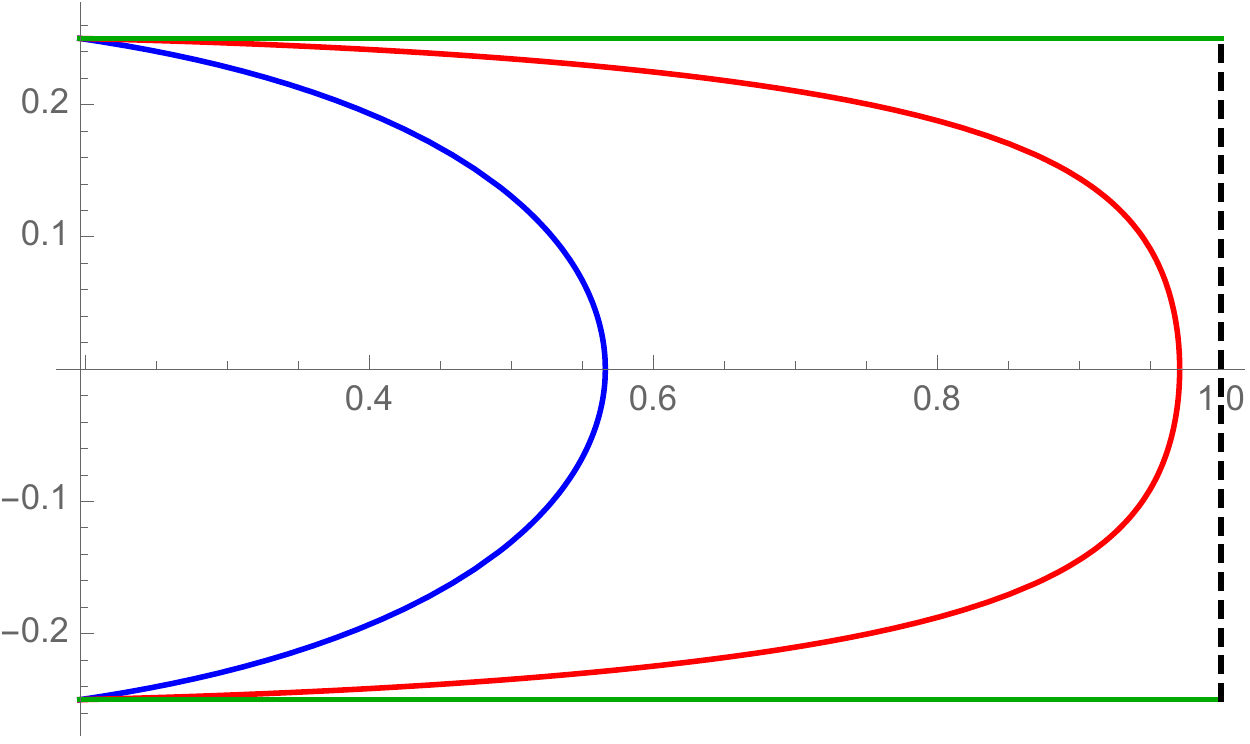}
\begin{picture}(0.1,0.1)(0,0)
\put(-252,120){\makebox(0,0){$mx/c$}}
\put(25,65){\makebox(0,0){$mr/c$}}
\end{picture}
\caption{Types of  geodesic for $J_2=1/2$, $J_1=1/5$.  The blue and red curves are the two possible connected geodesics and the pair of green lines is the disconnected geodesic.  All three geodesics end at $mr_\textrm{UV}/c=1/\sqrt{26}$ and the dashed line is the IR cut-off $mr_\textrm{IR}/c=1$. \label{fig:CFTCurvePlot}}
\end{center}
\vskip-1.5em
\end{figure}
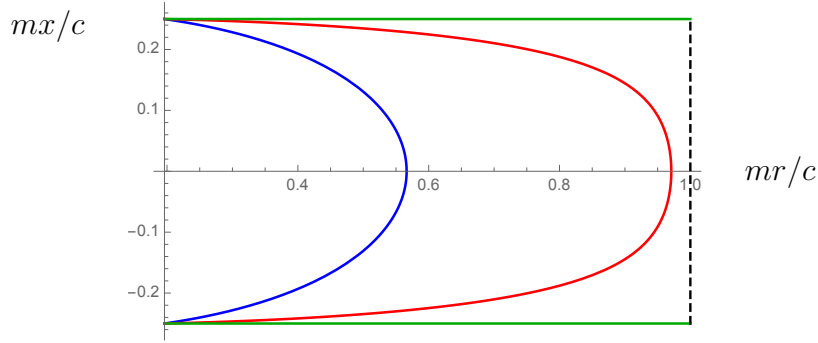

Two types of geodesic are relevant for a non-zero mass. See Figure~\ref{fig:CFTCurvePlot} for examples. The first type connects the  endpoints and is smooth at  the point of deepest extent in $r$: $x(r_\star)=0$ and $x'(r_\star)\to\infty$. The boundary condition for this type of geodesic relates $r_\star$ to the interval width~$\ell$:
\begin{equation}\label{eq:MassiveRelativisticStrip}
J_2 =\sqrt{1 - a^2}\, \tanh^{-1}\sqrt{a^2(1+J_1^2)-J_1^2}\ , \quad \quad a\equiv\frac{mr_\star}{c}\ .
\end{equation}
At fixed $J_1$, the function on the right-hand side is real and non-zero for $J_1(1+J_1^2)^{-1/2}< a< 1$, in accordance with \eqref{eq:CoordinateLimits}.  It is positive within this range and has a single maximum.  Thus,   equation~\eqref{eq:MassiveRelativisticStrip} cannot be satisfied for large $J_2$, but may have two solutions for small $J_2$. The second type of geodesic exists for all $J_2$ and consists of two disconnected straight sections $x(r) \equiv \pm\ell/2$ that end at $r = r_{\text{IR}}$.  

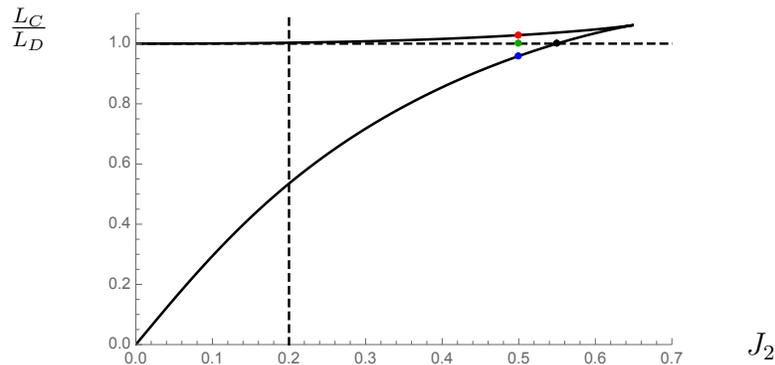
\begin{figure}[!h]
\vskip0.5em
\begin{center}
\includegraphics[width=0.45\textwidth]{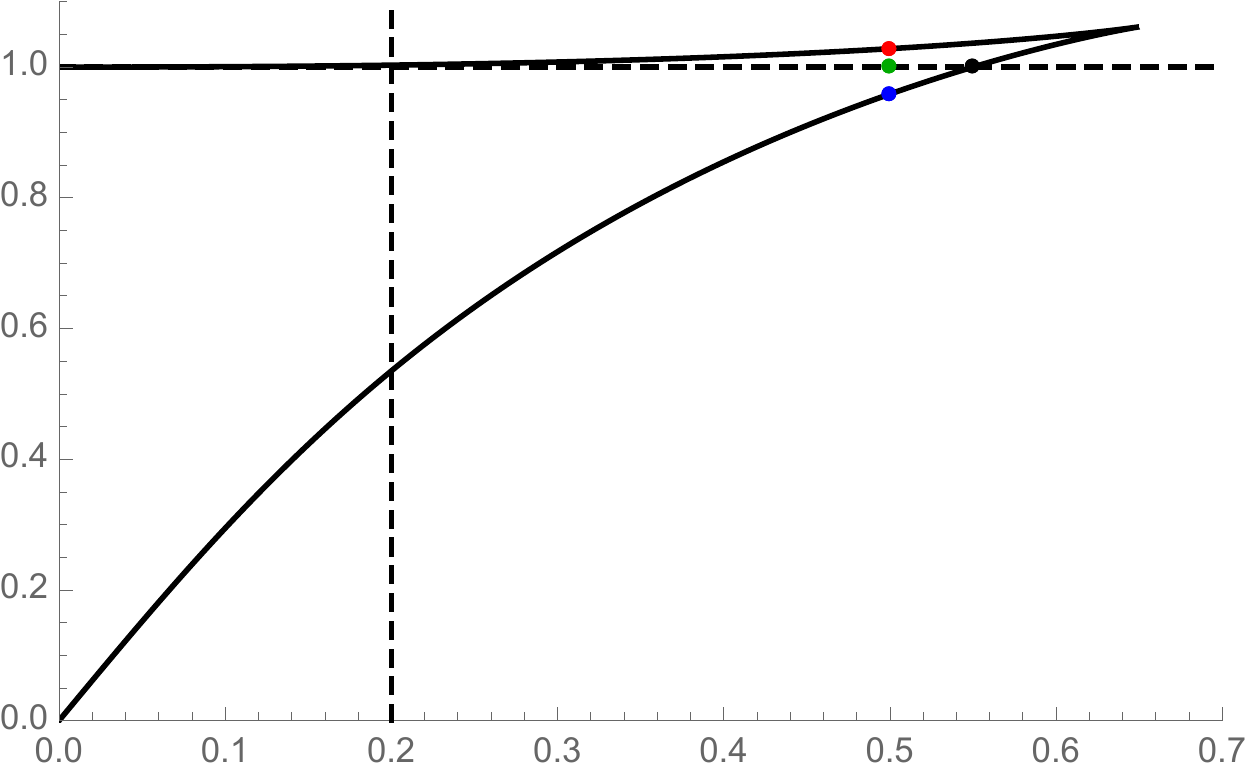}
\begin{picture}(0.1,0.1)(0,0)
\put(-252,130){\makebox(0,0){$\frac{L_C}{L_D}$}}
\put(25,10){\makebox(0,0){$J_2$}}
\end{picture}
\caption{Ratio of the connected geodesic length to the disconnected geodesic length for $J_1=1/5$ as a function of $J_2$.  
 The red,   green and blue  dots correspond to the curves plotted in Figure~\ref{fig:CFTCurvePlot} for $J_2=1/2$ and the black dot marks the critical value of $J_2$ for this $J_1$. The region to the left of the vertical dashed line is unphysical since $J_2<J_1$ therein. \label{fig:CFTLengthPlot}}
\end{center}
\vskip-1.5em
\end{figure}

We must be careful to identify the geodesic with the shortest length for a given $J_2$.  The lengths of the connected and disconnected geodesics are given respectively by 
\begin{align}
L_C &= \tanh^{-1}\left(\frac{\sqrt{a^2(1+J_1^2)-J_1^2}}{a}\right)-a \tanh^{-1}\sqrt{a^2(1+J_1^2)-J_1^2}\label{eq:MassiveRelativisticLength}\ ,\\
L_D & = \frac{1}{2}\log \left( 1 + \frac{1}{J_1^{2}}\right)\ . \label{eq:MassiveRelativisticLengthDisconnected}
\end{align}
We demonstrate in Figure~\ref{fig:CFTLengthPlot}  that the disconnected geodesic is shorter than any connected geodesic above a critical value of $J_2$. This value is slightly below that for which a connected geodesic no longer exists. 

In the massless case (i.e.\ setting $m=0$ from the beginning in \eqref{eq:CFTMetric}) the connected geodesic is always the shortest. We can find its length explicitly as a function of $\ell/\varepsilon$:
\begin{equation}\label{eq:MasslessRelativisticResult}
L = \log\left[ \frac{\ell}{\varepsilon}+\sqrt{\left(\frac{\ell}{\varepsilon}\right)^2+1} \right]=\log\left(\frac{\ell}{\varepsilon}\right)+\log 2 +\mathcal{O}\left(\frac{\varepsilon}{\ell}\right)^2\ .
\end{equation}
In the last equation, we made an expansion in small $\varepsilon/\ell$. The first term is the dominant term and leads to the area law, which in $1+1$ dimensions is logarithmic in $\ell$. The entanglement entropy is proportional to the length of the geodesic and the proportionality factor is known to be related to the central charge \cite{Holzhey:1994we}:
\begin{equation}
S=\frac{c}{3}\log\left(\frac{\ell}{\varepsilon}\right)+\cdots\ ,
\end{equation}
with $c=1$ for a real scalar field. 

Given that we have fixed the overall normalisation, we can return to the massive case and perform another consistency check for the case when the correlation length is smaller than the size of the interval. This translates into large $J_2$, for which the disconnected bulk curve \eqref{eq:MassiveRelativisticLengthDisconnected} is the shortest.  Expanding for small $J_1$ and using the same normalisation factor $c/3$ as before, we find the following leading term:
\begin{equation}\label{EE-CC}
S=\frac{c}{6}\log\left(\frac{1}{J_1^2}\right)+\cdots=\frac{c}{3}\log\left(\frac{\xi}{\varepsilon}\right)+\cdots\ ,
\end{equation}
where we recall that the correlation length is defined by $\xi\equiv c/m$. This result for the entanglement entropy matches perfectly with the universal result of Cardy and Calabrese in \cite{Calabrese:2004eu} applied to the case of one interval with two boundary points. 

In the opposite limit, i.e. for large correlation lengths compared to the length of the line interval, the cMERA approach seems to not reproduce known results. Namely, it was shown in \cite{Casini:2005zv,Casini:2009sr} that the entanglement entropy for a free massive scalar field in 1+1 dimensions in the regime $\ell\ll \xi$ contains a term proportional to $\log(-\log(m\varepsilon/c))=\log(-\log J_1)$ in the limit of small $J_1$. It is unclear to us how such a double logarithm would appear in the cMERA approach (see also Section 3.3 for the case of general $z$), and it would be interesting to understand better where this apparent discrepancy comes from. 

This concludes the consistency checks on the holographic cMERA techniques applied to free relativistic scalar field theories. The generalisation to higher dimensions can also be done, but we include it in the general analysis of Lifshitz theories with arbitrary dynamical exponent $z$. When we set $z=1$ we recover the relativistic results.

\section{Lifshitz entanglement entropy}\label{sec:Lifshitz}

We now apply the holographic cMERA technique to compute entanglement entropy in free Lifshitz scalar field theories. This possibility was in fact already pointed out in \cite{Nozaki:2012zj} but not worked out. In fact, in that reference, the case $z=1$ treated in the previous section was also not worked out in detail. The case of $d=1$ was recently also considered in \cite{He:2017wla} which we review and extend below. 

As suggested in \cite{Nozaki:2012zj}, all we need is the dispersion relation and the rest is computational. In this section we perform all the calculations and provide results for arbitrary $z$ and $d$, treat the massive case as well, and compute the RG flow of the entanglement entropy from $z>1$ theories to $z=1$.

\subsection{Massless Lifshitz scalar}

First we consider the simple theory given in \eqref{eq:LifshitzTheory} with $d=1$.  This has dispersion relation $\omega(k) = \alpha k^z.$  According to the  prescription given in the previous section, we should associate the following metric with its ground state:
\begin{equation}\label{eq:MasslessLifshitzMetric}
{\rm d}s^2 = \frac{z^2}{4}\, {\rm d}u^2 + \frac{e^{2u}}{\varepsilon^2}\, {\rm d}x^2+g_{tt}\, {\rm d}t^2\ .
\end{equation}
We wish to compute the entanglement entropy of an interval in this state, and for this we calculate the length of the geodesics in this metric.  Remarkably, the result is a simple rescaling of the massless relativistic result~\eqref{eq:MasslessRelativisticResult}:
\begin{equation}\label{eq:MasslessLifshitzResult}
L = z \log\left( \frac{\ell}{z\varepsilon}+\sqrt{\left(\frac{\ell}{z\varepsilon}\right)^2+1} \right)\ .
\end{equation}
This can be seen immediately at the level of the metric.  Suppose we scale out the factor $z^2$ in front of the metric. Then up to this overall rescaling, this is the $z=1$ metric provided we replace the cutoff $\varepsilon\to z \varepsilon$ in \eqref{eq:MasslessLifshitzMetric}. (Physically, we are not changing the cutoff $\varepsilon$ for the Lifshitz theory, it is just a trick to get the answer. Equivalently, one can also rescale $x\to z x$.) It is then clear that the length of geodesics on the $t=0$ slice is just a rescaling by $z$ of that obtained from~\eqref{eq:CFTMetric} at $m=0$. This observation was already made in \cite{He:2017wla}, but there the result was only given for large $\ell/z\varepsilon$, which does not  permit taking large values of $z$.

Here, we also obtain a sensible large $z$ limit.  In a discrete model of a continuum local  theory, entanglement entropy is dominated by contributions from nearest neighbors across the boundary of the entangling region.  As $z$ is increased, more and more lattice sites are involved and we expect the full interval to contribute at infinite $z$. This is indeed what we find:
\begin{equation}
\lim_{z\to\infty} L = \frac{\ell}{\varepsilon}\ ,
\end{equation}
i.e. it  becomes extensive with the region size, so we recover a volume law in the large $z$ limit.  This agrees with the observations and results of \cite{MohammadiMozaffar:2017nri}. It has to be noted though that this limit is only natural in the continuum limit (where also the cutoff $\varepsilon$ is sent to zero), as long as $z\gg l/\varepsilon$.

The entanglement entropy $S$ follows from $L$ by a multiplicative factor. In AdS/CFT, this factor comes from translating Newton's constant into the central charge of the theory, which in 1+1 dimensions gives a multiplicative factor of $c/3$, leading to $S=c/3 \log(\ell/\epsilon)+c_0$ for relativistic CFTs. The cMERA approach, however, does not fix the overall normalization. The best we can do is to replace the central charge $c$ by an overall multiplicative constant $c_z$, independent of $\ell$ but which can depend on the dynamical exponent $z$. For a relativistic scalar field, we have $c_{z=1}=1$. In general, we call $c_z$ the Lifshitz central charge. Then we obtain the following  entanglement entropy formula for general values of $z$:
\begin{equation}\label{EEanyz&l}
S_{d=1}=\frac{c_z}{3} z\log\left( \frac{\ell}{z\varepsilon}+\sqrt{\left(\frac{\ell}{z\varepsilon}\right)^2+1} \right)\ .
\end{equation}
There might still be a non-universal additive constant independent of $\ell$, just like the coefficient $c_0$ that is cutoff-dependent. We will leave it out of the discussion here.

The expansion around small values of $\ell/z\varepsilon$ gives small deviations from the volume law,
\begin{equation}
S_{d=1}=\frac{c_z}{3}\frac{\ell}{\varepsilon} \left[ 1-\frac16\left(\frac{\ell}{z\varepsilon}\right)^2+\mathcal{O}\left(\frac{\ell}{z\varepsilon}\right)^4\right]\ ,
\end{equation}
whereas expanding around the area law $\ell/z\varepsilon \gg 1$ gives deviations from the area law,
\begin{equation}
S_{d=1}=\frac{c_z}{3} z\left[ \log\left(\frac{2\ell}{z\varepsilon}\right)+\frac{z^2\epsilon^2}{4\ell^2}+\mathcal{O}\left(\frac{z\varepsilon}{\ell}\right)^4\right]\ .
\end{equation}
In \cite{He:2017wla}, the assumption was made that $c_z$ is independent of $z$. Recent numerical work, however, shows that this assumption should be relaxed, as the work of \cite{Chen:2017txi,Chen:2017tij,MohammadiMozaffar:2017chk} showed that for $z=2$, one has $c_{z=2}=3/4$, such that the entanglement entropy for $z=2$ starts like $S=\frac{1}{2}\log(\frac{\ell}{\epsilon})$ to leading order. This implies that $c_z$ does depend on $z$. But the fact that we don't know $c_z$ does not prevent us from making predictions that could be checked using numerics or other methods. Indeed for fixed but arbirtray $z$, we can still check the functional dependence on $\ell/z\epsilon$ inside the logarithm in \eqref{EEanyz&l} with numerical methods. Or stated differently, we can take the ratio of the entanglement entropies for large and small values of $\ell/z\epsilon$ and compare this to lattice results. This ratio is independent of $c_z$, so is not sensitive to our ignorance of it. Admittedly, this issue needs to be better understood, either from further numerical work, or perhaps analytically, by computing $c_z$ from the replica trick or from Lifshitz scale anomaly coefficients. We leave this for further study, and leave the coefficient $c_z$ undetermined in this paper. More conveniently, we will focus on just the length or area of the minimal surfaces, which do not involve this factor $c_z$.

We now generalise our treatment to higher dimensions.  We make the replacement ${\rm d}x^2\to {\rm d}\vec{x}^{\,2}_d$ with $\vec{x}=(x_1,...,x_d)$ and also the change of coordinates $e^{2u} = \frac{\varepsilon^2}{r^2}$ in \eqref{eq:MasslessLifshitzMetric}:
\begin{equation}
{\rm d}s^2=\frac{z^2}{4}\frac{{\rm d}r^2}{r^2}+\frac{{\rm d}\vec{x}^{\,2}_d}{r^2}+g_{tt}\,{\rm d}t^2\ .
\end{equation}
Our region of interest is now the strip $\left\{\vec{x} \left| -\frac{\ell}{2}\leq x_d \leq \frac{\ell}{2}\right.\right\}$ at $t=0$. Translational invariance of the metric and strip in the additional spatial directions simplifies the problem dramatically: we seek  extremal surfaces of this metric parametrised by $x_d(r)$ that fill the other directions and end at $x_d(\varepsilon)=\pm\ell/2$.    The area of the smallest surface is proportional to the entanglement entropy.

We find the following expressions for the width of the strip and the area of the surface extending to $r=r_\star$, respectively:
\begin{align}
\frac{\ell}{\varepsilon} &= z\, \frac{b}{2 d}\, \Gamma\left(\tfrac{1 + d}{2 d}\right) \left[\frac{\sqrt{\pi}}{\Gamma\left( \frac{1+2d}{2 d}\right)} - \frac{ _2F_1\left(\frac{1}{2}, \frac{1 + d}{2 d}, \frac{1+3d}{2d}, 
     \frac{1}{b^{2 d}}\right)}{\Gamma\left(\frac{1+3d}{2d}\right)\, b^{d+1}} \right] \ ,\label{eq:ellMasslessLifshitz}\\
 A    &= z\, \frac{V_{d-1}}{\varepsilon^{d-1}} \left[\frac{\sqrt{\pi}\, \Gamma\left( \frac{1-d}{2 d}\right)}{2d\, \Gamma\left( \frac{1}{2 d}\right)\, b^{d-1}} - \frac{ _2F_1\left(\frac{1}{2}, \frac{1 - d}{2 d}, \frac{1+d}{2d}, 
     \frac{1}{b^{2 d}}\right)}{d-1} \right]\ ,\label{eq:LengthMasslessLifshitz}
\end{align}
where we have defined $b\equiv r_\star/\varepsilon$ and regulated the infinite volume of the remaining directions with $V_{d-1}$. Note  that these results are again related to those of the relativistic case with a rescaling by $z$: we scale $\varepsilon\rightarrow z\varepsilon$ and multiply $A\rightarrow z^d A$ .  

We deduce from these results that the entanglement entropy follows an area law for finite~$z$.  The equation~\eqref{eq:ellMasslessLifshitz} relating $\ell$ and $r_\star$ has a unique solution for given $d$ and $z$.  We can invert this asymptotically at large $\ell/\varepsilon$ and find the following expansion for the area: 
\begin{equation}
A = \frac{z}{d-1}\, \frac{V_{d-1}}{\varepsilon^{d-1}}  - \frac{(z\kappa)^d}{d-1}\, \frac{V_{d-1}}{\ell^{d-1}} + O\left(\frac{V_{d-1}\,\varepsilon^{d+1}}{\ell^{2d}}\right), \quad \quad \kappa(d) \equiv \sqrt{\pi}\, \frac{\Gamma\left(\frac{d+1}{2d}\right)}{\Gamma\left(\frac{1}{2d}\right)}\ .
\end{equation}
The first term represents the area term, and the second term is finite and independent of the cutoff. (See also, for example,  \cite{Ryu:2006ef} for the case $z=1$ using AdS/CFT and the RT formula.) The entanglement entropy is proportional to $A$, with a proportionality factor that is not universal  for $d>1$. In the cMERA approach, this normalisation inherits from that of the disentangler operator ${\cal K}(u)$ and is usually fixed by comparing with known field theory results for the entanglement entropy. Because of the non-universality, we leave this overall factor undetermined. 

In the large $z$ limit, we again obtain a volume law instead of an area law.  We can invert~\eqref{eq:ellMasslessLifshitz} asymptotically at large $z$, finding
\begin{equation}
b = 1 + \frac{1}{2} \left( \frac{\sqrt{d}\ell}{z\varepsilon} \right)^2 + \frac{2d-5}{24} \left( \frac{\sqrt{d}\ell}{z\varepsilon} \right)^4 + \ldots\ .
\end{equation}
We substitute this into the area expression \eqref{eq:LengthMasslessLifshitz} and find
\begin{equation}
\lim_{z\to\infty}A = \frac{\ell\, V_{d-1}}{\varepsilon^d}\ , 
\end{equation}
which is indeed proportional to the regulated volume $\ell\, V_{d-1}$ of the strip.

\subsection{Disc geometry}

In this section we  depart briefly from the strip to consider a region of finite size: the disc.  We study the same  state of theory~\eqref{eq:LifshitzTheory} but simply write the appropriate cMERA metric in different coordinates:
\begin{equation}\label{eq:MasslessLifshitzMetricDisc}
{\rm d}s^2 = \frac{z^2}{4} \, \frac{{\rm d}r^2}{r^2} +  \frac{{\rm d}p^2+p^2\, {\rm d}s^2_{S^{d-1}}}{r^2} +g_{tt}\, {\rm d}t^2\ .
\end{equation}
The following function describes a surface lying on the constant time slice $t=0$ that ends at $r=\varepsilon$ on a disc of radius $p=R$:
\begin{equation}
r(p)=\varepsilon\left(\frac{2R}{z\varepsilon}\right)\sqrt{1- \frac{p^2}{R^2} + \left(\frac{z \varepsilon}{2R}\right)^2}\ .
\end{equation}
This surface is smooth at $p=0$, independent of $d$, minimises the area functional and has area
\begin{equation}\label{EEdisc}
A = \textrm{Vol}\left(S^{d-1}\right) \frac{z^d}{2^d d}\left(1+x^2\right)^{-d/2}\, _2F_1\left(\frac{d}{2},\frac{d+1}{2};\frac{d+2}{2};\frac{1}{1+x^2}\right), \quad \quad x\equiv \frac{z \varepsilon}{2R}\ .
\end{equation}
The area diverges as the cut-off is taken much smaller than the disc radius.  For finite $z$, this divergence is proportional to the area of the disc:
\begin{equation}
A=\frac{z}{2(d-1)}\, \frac{\textrm{Vol}\left(S_R^{d-1}\right)}{\varepsilon^{d-1}}+ \ldots\ .
\end{equation}
The lowest subleading divergence is logarithmic when $d$ is odd; in particular, we recover~\eqref{eq:MasslessLifshitzResult} for $d=1$  since the  interval width satisfies $\ell = 2R$.  For even $d$, the small $x$ expansion contains a constant term which is universal. All these results are straightforward generalisations and rescalings of the $z=1$ case described  in, for example, \cite{Ryu:2006ef}. 

It is interesting to compare our results with \cite{Fradkin:2006mb}, where the entanglement entropy was studied for $d=2$ with $z=2$. In particular it was found that for the disc subspace, no logarithmic terms were found, see case (a) in Figure 1 in \cite{Fradkin:2006mb}. This is consistent with our general result that for even $d$, no logarithms appear in the expansion of \eqref{EEdisc}. More interesting would be to study the geometries where logarithmic terms do arise, such as the rectangular or half disc subspaces studied in \cite{Fradkin:2006mb} as well. To reproduce the coefficients in front of the log-terms using the cMERA approach, we would need to perform extremisation of surfaces in geometries that end on these rectangles or half discs that have less symmetry. This is much harder in $d\geq 2$ but is an interesting problem for separate study. Given that we don't know proportionality coefficient between the area and the entanglement entropy, it is cumbersome to perform such a test at present.

In the large $z$ limit we again obtain a volume law:
\begin{equation}
\lim_{z\to\infty}A= \frac{\textrm{Vol}\left(S^{d-1}\right)}{d}\left(\frac{R}{\varepsilon}\right)^d=\frac{\textrm{Vol}\left(B^{d}_R\right)}{\varepsilon^d}\ ,
\end{equation}
Here $B_R^d$ is the $d$-dimensional ball whose boundary is the $(d-1)$-dimensional sphere of radius~$R$. 

From now on we focus exclusively on strip regions.

\subsection{Massive deformation}

We now deform the theory  \eqref{eq:LifshitzTheory} by adding the mass term $m^2\phi^2$.  The dispersion relation is modified to $\omega(k)^2 = \alpha^2k^{2z}+m^2$, leading to a cMERA metric 
\begin{equation}\label{eq:MassiveLifshitzMetric}
{\rm d}s^2 = \left[\frac{z\, e^{2zu}}{2\left(e^{2zu}+(m\varepsilon^z/\alpha)^2\right)}\right]^2 {\rm d}u^2 + \frac{e^{2u}}{\varepsilon^2}\, {\rm d}\vec{x}^{\,2}_d+g_{tt}\, {\rm d}t^2\ .
\end{equation}
The dimensionless parameters that characterise the mass deformation and the region size can be chosen  
\begin{equation}
J_1\equiv \frac{\varepsilon}{\xi} \quad \text{and} \quad J_2\equiv \frac{\ell}{\xi}\ ,\qquad \xi\equiv (\alpha/m)^{1/z}\ ,
\end{equation}
respectively. They reduce to \eqref{eq:J1andJ2Relativistic}  in the relativistic limit.  We have written them in terms of the correlation length $\xi$, generalising the one from the relativistic case.
We want to compute entanglement entropy as a function of $d$, $z$, $J_1$ and $J_2$.  Again we require $J_1<1$ and $J_2 > J_1 $  to ensure that the cut-off $\varepsilon$ is the smallest length scale in the theory. As in the relativistic case, the correlation length maybe larger or smaller than the interval length or strip width~$\ell$, corresponding to the regimes $J_2<1$ or $J_2>1$ respectively.

Just like the relativistic case covered in Section~\ref{sec:Consistency}, there is a competition between  two types of extremal surface: connected and disconnected.  We analyse this case in a similar fashion. We begin with a change of coordinates:
\begin{equation}
e^{2zu}=\frac{\varepsilon^{2z}}{r^{2z}}-J_1^{2z} \quad \Longrightarrow \quad {\rm d}s^2 =  \frac{z^2{\rm d}r^2}{4r^2} + \left(\frac{\varepsilon^{2z}}{r^{2z}}-J_1^{2z}\right)^{1/z}\frac{{\rm d}\vec{x}^{\,2}_d}{\varepsilon^2}+g_{tt}\, {\rm d}t^2\ .
\end{equation}
For a connected surface extending to $r=r_\star$, the width of the strip and the area of the surface can be written
\begin{align}
J_2 &= z \int_{J_1}^{t_\star} {\rm d}t\, \frac{(t/t_\star)^d}{(1+t^{2z})\sqrt{1-(t/t_\star)^{2d}}} \ ,\label{eq:ellMassiveLifshitz}\\
A_C &= z\, J_1^{(d-1)}\, \frac{V_{d-1}}{\varepsilon^{d-1}} \int_{J_1}^{t_\star} {\rm d}t\, \frac{1}{t^d\, (1+t^{2z})\sqrt{1-(t/t_\star)^{2d}}} \ ,\label{eq:AreaMassiveLifshitz}
\end{align}
respectively, where it is convenient to parametrise the bulk depth via 
\begin{equation}
a\equiv\frac{r_\star}{\xi} \quad \textrm{with}\quad   t_\star\equiv\left(\frac{a^{2z}}{1-a^{2z}}\right)^{\frac{1}{2z}}\ .
\end{equation}
We can  evaluate these integrals exactly for various values of $d$ and $z$; more generally, they are straightforward to evaluate numerically.   Equation~\eqref{eq:ellMassiveLifshitz} relates $J_2$ and $r_\star$.  We find that  it may have zero, one or two solutions depending on the value of $J_2$, just like the relativistic case.  We observe an area law divergence in $A_C$ that degenerates to a logarithmic divergence for $d=1$.  

For example, we have obtained analytical expressions for $d=1$, $z\in \mathbb{N}$:
\begin{align}
J_2 &= \frac{1}{t_\star}\sum_{n=1}^z\frac{\gamma_n}{\sqrt{1+\gamma_n/t_\star^{2}}}\, \tanh^{-1}\sqrt{\frac{1-J_1^2/t_\star^{2}}{1+\gamma_n/t_\star^{2}}} \ ,\\
L_C &=  z\tanh^{-1}\sqrt{1-J_1^2/t_\star^{2} }-\sum_{n=1}^z\frac{1}{\sqrt{1+\gamma_n/t_\star^{2}}}\, \tanh^{-1}\sqrt{\frac{1-J_1^2/t_\star^{2}}{1+\gamma_n/t_\star^{2}}}\ , \label{eq:MassiveLifshitzLengthAnalytical}
\end{align}
where the $\gamma_n$ satisfy 
\begin{equation}
1+(-1)^z\, \gamma_n^z = 0\quad \Longrightarrow \quad \gamma_n = e^{ i \pi (2n-1+z)/z},\quad  n=1,2,\ldots,z\ .
\end{equation}
It is clear from the first term in \eqref{eq:MassiveLifshitzLengthAnalytical} that the length diverges like $-\log J_1$ as $J_1\to 0$.  Note  that the correct formulae \eqref{eq:MassiveRelativisticStrip} and \eqref{eq:MassiveRelativisticLength} are  recovered for $z=1$.  In addition, analytical expressions can be obtained for $z=d$, but these are lengthy and not  very illuminating.

The disconnected surface exists for all $J_2$. Its area is in fact independent of $J_2$ and can be evaluated in closed form:
\begin{align}
A_D &= z\, J_1^{(d-1)}\, \frac{V_{d-1}}{\varepsilon^{d-1}} \int_{J_1}^{\infty} {\rm d}t\, \frac{1}{t^d\, (1+t^{2z})}\nonumber \\
 &= \frac{z}{d-1}\, \frac{V_{d-1}}{ \varepsilon^{d-1}}\left[1- {_2F_1}\left(1,\frac{d-1}{2z};\frac{d-1}{2z}+1;-\frac{1}{J_1^{2z}}\right)\right], \quad \quad d>1
\end{align}
This result is again proportional to the area of the region. 
For $d=1$ we find a simple expression that diverges logarithmically as $J_1\to 0$:
\begin{equation}\label{smallJ1LD}
L_D = \frac{1}{2}\log \left( 1 + \frac{1}{J_1^{2z}}\right)\ .
\end{equation}

We must now identify which surface has the least area for a given region size $J_2$.  For fixed $d$, $z$ and $J_1$ we find that the disconnected surface has minimal area above a critical value of $J_2$. This value is slightly below that for which a connected surface no longer exists.  
We observe the same qualitative behaviour as presented in Figure~\ref{fig:CFTLengthPlot} for the relativistic case regardless of $d$ and $z$.  Besides yielding analytical solutions, the point $z=d$ does not appear to be special in this calculation.

\begin{figure}[!h]
\begin{center}
\includegraphics[width=0.45\textwidth]{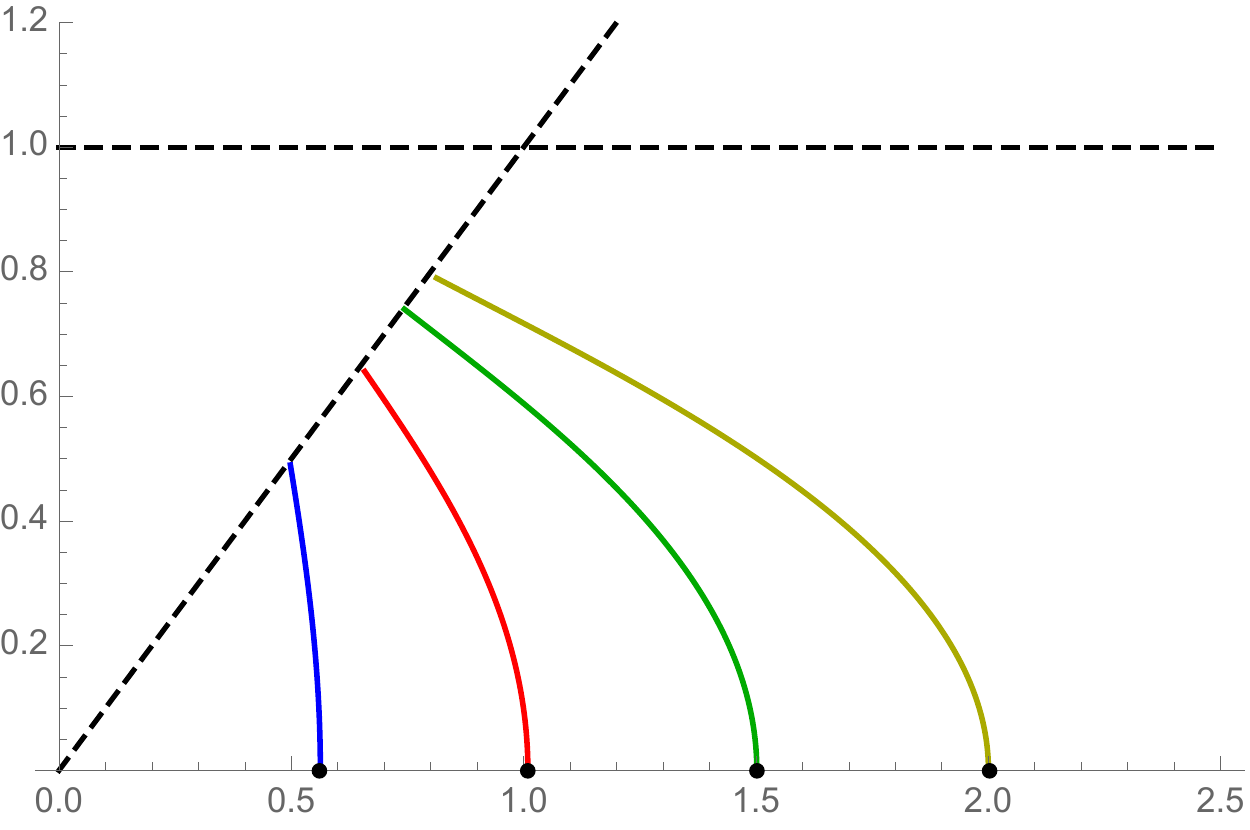}
\begin{picture}(0.1,0.1)(0,0)
\put(-252,130){\makebox(0,0){ $J_1$}}
\put(25,10){\makebox(0,0){$J_2$}}
\end{picture}
\caption{Phase diagram for $d=1$.  The curves correspond to  $z=1$ (blue), $z=2$ (red), $z=3$ (green) and $z=4$ (yellow).  To the left of each curve the shortest geodesic is  connected,  whereas to the right  the shortest  geodesic is disconnected. The dashed lines mark the boundaries of the physical region $J_1<1$ and $J_2>J_1$. The black dots are extracted directly at $J_1=0$.\label{fig:PhaseDiagram}}
\end{center}
\vskip-1.5em
\end{figure}
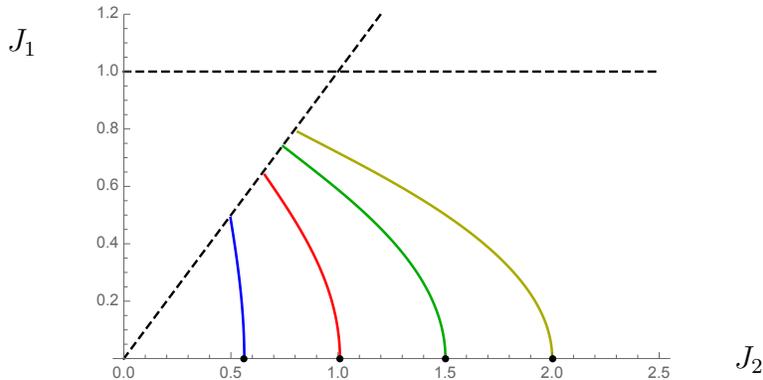

For any given $J_1$ it is   straightforward  to numerically find the critical value of $J_2$ at which the two types of surface have equal area.  In this way we can construct the phase diagram of a given theory.  Our results in $d=1$ for various values of $z$ are plotted in  Figure~\ref{fig:PhaseDiagram}.  Note that the $J_2$-intercepts  of the  phase boundaries follow a roughly linear relationship: $J_2(J_1= 0)\sim z/2$.  
For large values of $J_2$  the disconnected curve is always the shortest. This is the limit in which the correlation length is smaller than the length of the interval: $\varepsilon \ll \xi \ll \ell$. For $z=1$ this was the regime in which the result of Cardy and Calabrese holds: c.f.\ \eqref{EE-CC}. For $z>1$, the result obtained in \eqref{smallJ1LD} generalises the Cardy-Calabrese result, and we obtain
\begin{equation}
S_{d=1}=\frac{c_z}{3}z\log \left(\frac{\xi}{\varepsilon}\right)\ ,
\end{equation}
where we have used the same normalisation factor $c_z/3$ as before that relates the length of the geodesic to the entanglement entropy for a real scalar field. 

For large correlation lengths $\xi> \ell$, so small $J_2$ (and therefore small $J_1$), we get again logarithmically diverging terms $-\log J_1$. There are no known analytical or numerical results in this case, but we expect similar discrepancies as for $z=1$, where the leading diverging term involves a double logarithm, $\log(-\log J_1)$ in the limit of small $J_1$.

\subsection{RG flow to an IR CFT} 
In this section we begin with a massless free scalar field in $1+1$ dimensions and turn on an irrelevant Lifshitz coupling:
\begin{equation}
I =\frac{1}{2} \int {\rm d}^{2}x\left[\left(\partial_t\phi\right)^2-c^2(\partial_x\phi)^2-\alpha^2(\partial_x^z\phi)^2\right]\ .
\end{equation}
Our goal is to compute the entanglement entropy along the entire renormalisation group flow from the UV Lifshitz theory to the IR CFT.  For sublattice entanglement entropy, such a study was carried out in \cite{He:2017wla}, where it was found that the entanglement entropy decreases along the renormalisation group flow, for any starting value of $z>1$.  This provides evidence for a generalisation of the c-theorem for entanglement entropy in the relativistic case \cite{Casini:2004bw}, applicable to flows between two Lifshitz fixed points. Our analysis below will give further support for this.

The dispersion relation is given by $\omega(k)^2=c^2k^2+\alpha^2 k^{2z}$.  After a change of coordinates, this results in a cMERA metric of the following form:
\begin{equation}
{\rm d}s^2   =  \frac{f(r)^2}{4 r^2}\, {\rm d}r^2 +  \frac{{\rm d}x^2}{r^2}   + g_{tt}\, {\rm d}t^2\ . \\
\end{equation}
The function $f(r)$ interpolates between the two limits of the flow as the dimensionless parameter $K\equiv\alpha/(c\, \varepsilon^{z-1})$ is varied:
\begin{equation}
f(r) = \frac{1+z K^2 (\varepsilon/r)^{2(z-1)}}{1+ K^2 (\varepsilon/r)^{2(z-1)}}\longrightarrow \left\{ 
\begin{array}{c c}
 z, & K \gg 1 \\
1, & K \ll 1
\end{array}
 \right.
\end{equation}
We seek geodesics of the form $x(r)$ extending to $r=r_\star$ that end at $x(\varepsilon)=\pm\ell/2$.  

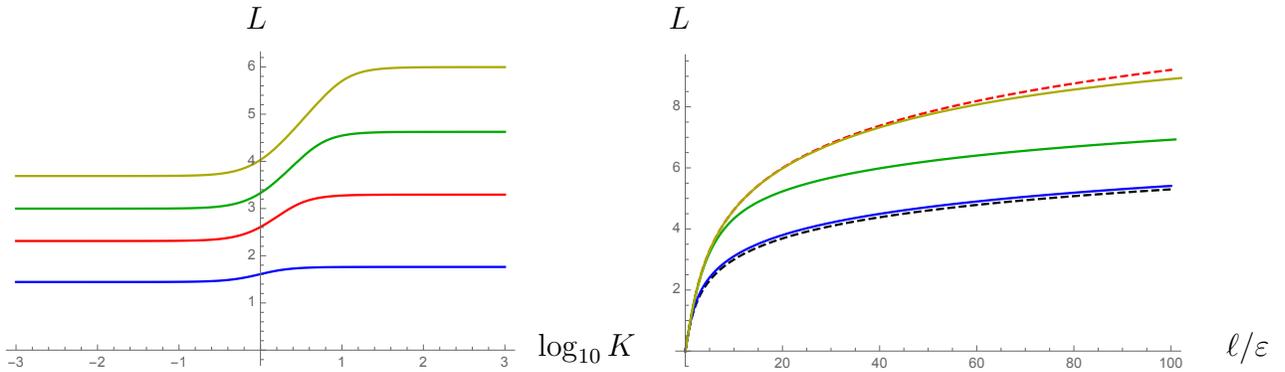
\begin{figure}[!h]
\vskip1.5em
\begin{center}
\hskip-2.5em
\includegraphics[width=0.4\textwidth]{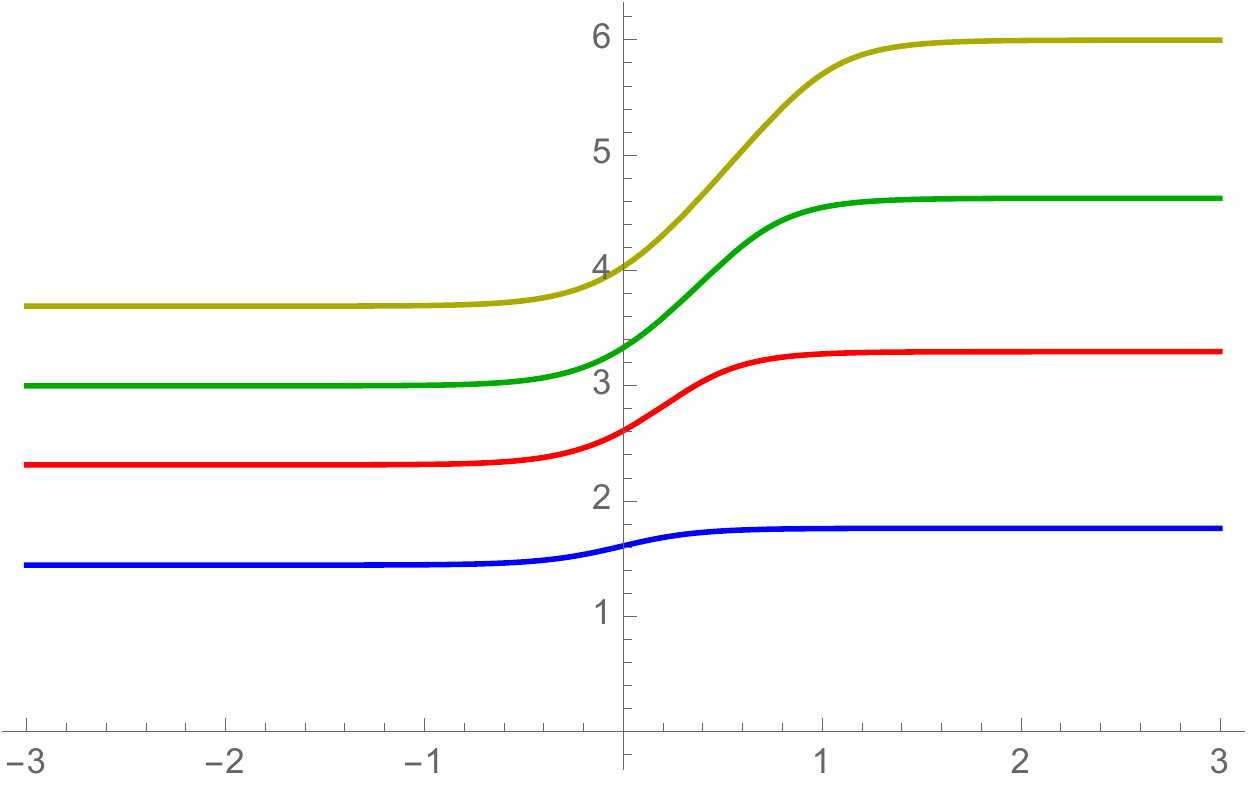}\hskip5em 
\includegraphics[width=0.4\textwidth]{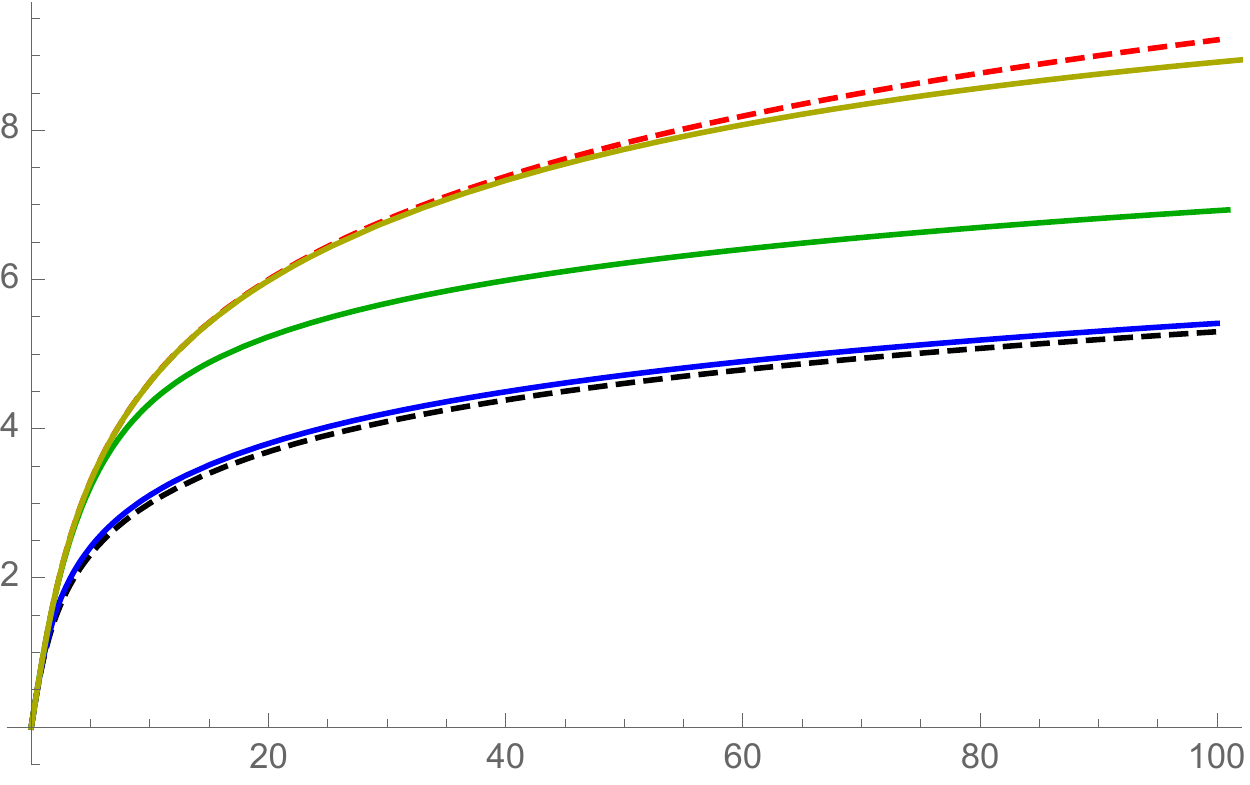}
\begin{picture}(0.1,0.1)(0,0)
\put(-355,135){\makebox(0,0){$L$}}
\put(-195,135){\makebox(0,0){$L$}}
\put(20,10){\makebox(0,0){$\ell/\varepsilon$}}
\put(-230,10){\makebox(0,0){$\log_{10}K$}}
\end{picture}
\caption{Entanglement entropy for a flow between a CFT in the IR and $z=2$ Lifshitz theory in the UV. Left: Length as a function of $\log_{10}K$ for $\ell/\varepsilon=2,5,10,20$ (bottom to top). These curves interpolate monotonically between the correct IR (left) and UV (right) limits. Right: Length as a function of $\ell/\varepsilon$ for $K=0.5,5,50$ (bottom to top, solid). The dashed lines correspond to the IR (black) and UV (red) limits. \label{fig:RG}}
\end{center}
\vskip-2em
\end{figure}

For example, for $z=2$ we find that the interval width and the geodesic length can be written respectively as
\begin{align}
\frac{\ell}{\varepsilon} &= \sqrt{b^2-1} + \frac{K^2}{\sqrt{b^2+K^2}}\, \tanh^{-1}\sqrt{\frac{b^2-1}{b^2+K^2}} \ ,\\
 L&= 2 \log\left(b+\sqrt{b^2-1}\right) - \frac{b}{\sqrt{b^2+K^2}}\, \tanh^{-1}\sqrt{\frac{b^2-1}{b^2+K^2}}\ ,
\end{align}
where we have again defined $b\equiv r_\star/\varepsilon$.    Note that the expression for $\ell/\varepsilon$ increases monotonically with $b$ (at fixed $K$) so can be inverted uniquely.   It is straightforward to check that these results have the correct limits: \eqref{eq:MasslessLifshitzResult} for large $K$ and \eqref{eq:MasslessRelativisticResult} for small $K$.  These are the UV and IR limits, respectively.
We  plot the geodesic length as a function of $K$ or $\ell/\varepsilon$ in Figure~\ref{fig:RG}.  We see that the length decreases monotonically along this RG flow from UV to IR, as expected. (We have checked that qualitatively similar results can be obtained numerically for other values of~$z$.)

\section{Discussion}\label{sec:Discussion}

An approach was put forward in \cite{Nozaki:2012zj} to geometrise entanglement entropy based on cMERA techniques \cite{Haegeman:2011uy}.  It is similar in spirit to AdS/CFT but is applicable to free fields, i.e.\ weak coupling and small $N$. We have illustrated in this paper that this approach leads to concrete predictions for the entanglement entropy of free scalar fields. In the relativistic case, it reproduces well-known analytical field theory results, as we showed in our analysis.\footnote{With the exception of the massive case where the correlation length exceeds the length of the interval, as discussed in Section 2.2.} In the Lifshitz case, it leads to new predictions, generalising and extending the observations made in \cite{He:2017wla}. The overall normalisation for the entanglement entropy is however not fixed from the metrics introduced by \cite{Nozaki:2012zj}; rather, it depends on the choice and normalisation of the disentangler operator. In AdS/CFT this normalisation is fixed by the dictionary that relates Newton's coupling constant to the central charge or number of colours. We have not been able to determine this overall normalization, which we denoted by $c_z$, and which can depend on the value of the dynamical exponent $z$. It would be interesting to understand if $c_z$ is related to Lifshitz scale anomaly coefficients. These anomalies have been studied in \cite{Arav:2014goa,Pal:2016rpz,Arav:2016akx}, but also in earlier work for $z=2$ \cite{Griffin:2011xs,Baggio:2011ha} in 2+1 and 3+1 dimensions, and $z=3$ in 3+1 dimensions \cite{Adam:2009gq}. The results obtained in e.g. \cite{Fradkin:2006mb} show that central charges do appear in certain logarithmic terms entanglement entropy for $z=d=2$ (so 2+1 dimensions), but it is not straightforward to see the relation to Lifshitz scale anomaly coefficients. Moreover, for the strip and disc geometries we consider here, these logarithmic terms are absent. It would be important progress to find such a relation if it exists.

In the massless case, our results for free Lifshitz scalars are obtained from a simple rescaling of the relativistic case, but this is not true in the massive case. In the latter case, we generalised the Calabrese-Cardy result to Lifshitz scalar fields with values of the dynamical exponent $z>1$. It would be interesting to repeat our analysis for fermions. 

Clearly, it would be important to better understand the validity of the holographic cMERA approach. From this point of view, our results are merely predictions rather than solid results. It would be nice to reproduce some of our predictions for Lifshitz entanglement based solely on field theoretic techniques. For free Lifshitz scalar fields, one imagines this should not be so difficult by attempting the replica method, but the presence of longer range interactions for large $z$ complicates this. In low dimensions, numerical and lattice methods can also be used, as was illustrated in \cite{MohammadiMozaffar:2017nri,He:2017wla}. We leave this topic for future study.

\section*{Acknowledgements}

It is our pleasure to thank Temple He, Javier Mag\'{a}n, Ali Mollabashi, Aurelio Romero-Berm\'{u}dez, Philippe Sabella-Garnier and William Witczak-Krempa for useful and enjoyable discussions and correspondence.  This work was supported in part by the Delta-Institute for Theoretical Physics (D-ITP) that is funded by the Dutch Ministry of Education, Culture and Science (OCW).

\providecommand{\href}[2]{#2}\begingroup\raggedright\endgroup


\end{document}